\documentclass{aa}
\usepackage{graphics}
\usepackage{rotating}   
\usepackage{psfig}
\usepackage{aabib99}
\usepackage{journals}
\newcommand{\gb}{\gamma_{\rm j}\beta_{\rm j}}
\newcommand{\figwidth}{0.48\textwidth}
\begin{document}

   \thesaurus{03 (     % A&A Section 03
              	10.03.1;	% Galaxy: center,
	      	11.01.2;  	% Galaxies: active,
               	11.10.1;  	% Galaxies: jets,
		13.25.2;	% X-rays: galaxies,
               	02.01.2;	% Accretion,	
		02.02.1	  	% Black hole physics
)}

   \title{The jet model for Sgr A*: radio and X-ray spectrum}

%   \subtitle{}

   \author{Heino Falcke
   \and    Sera Markoff$^1$}

   \titlerunning{Jet model for Sgr A*}
    \authorrunning{Falcke \& Markoff}
   \offprints{hfalcke@mpifr-bonn.mpg.de}

   \institute{Max-Planck-Institut f\"ur Radioastronomie,
              Auf dem H\"ugel 69, 53121 Bonn, Germany
		(hfalcke,smarkoff@mpifr-bonn.mpg.de)
             }

   \date{Astronomy \& Astrophysics, Vol. 362, p. 113 (2000)}

   \maketitle

   \begin{abstract} The preliminary detection of the Galactic Center
   black hole Sgr A* in X-rays by the Chandra mission, as well as
   recent mm-VLBI measurements, impose strict constraints on this
   source. Using a relativistic jet model for Sgr A*, we calculate the
   synchrotron and synchrotron self-Compton emission.  The predicted
   spectrum provides an excellent fit to the radio spectrum and the
   the X-ray observations.  Limits on the infrared flux and the low
   X-ray flux require a high-energy cut-off in the electron spectrum at
   $\gamma_{\rm e}\la100$. The implied lack of a significant power-law
   tail of high-energy electrons also suppresses the appearance of the
   extended, optically thin radio emission usually seen in
   astrophysical jets. The jet therefore appears rather compact and
   naturally satisfies current VLBI limits.  The initial parameters of
   the model are tightly constrained by the radio spectrum and the
   "submm-bump" in particular.  While the jet most likely is coupled
   to some kind of accretion flow, we suggest that the most visible
   signatures can be produced by this outflow.  If SSC emission from
   the jet contributes to the Sgr A* spectrum, significant variability
   in X-rays would be expected. The model could be generic for other
   low-luminosity AGN or even X-ray binaries.  \end{abstract}

%
%________________________________________________________________
\footnotetext[1]{Alexander-von-Humboldt research fellow}
\section{Introduction}
Sagittarius A* (Sgr A*) is the compact radio source at the Galactic
Center, which mounting evidence suggests is the signature emission of
a supermassive black hole.  Proper motion studies with radio
interferometers \cite{ReidReadheadVermeulen1999,BackerSramek1999a}
place Sgr A* at the apparent dynamical center of the Galaxy, with a
lower limit on its mass of $\sim 10^3M_{\sun}$.  Similarly the motions
of the surrounding stars, as well as near-infrared (NIR) line
spectroscopy, indicate a mass of $2.6\cdot10^6M_{\sun}$ enclosed
within $\sim 0.01$ pc
\cite{HallerRiekeRieke1996,EckartGenzel1996,GhezKleinMorris1998}.
These measurements rule out any known multiple star system.

At the same time, Very Long Baseline Interferometry (VLBI) places
strict upper limits on the frequency-dependent size and structure of
Sgr A*. Observations at 86 \& 220 GHz
\cite{RogersDoelemanWright1994,KrichbaumGrahamWitzel1998} constrain
Sgr A* to be around 0.06-0.2 milli-arcseconds (mas) at these
frequencies.

The radio spectrum of Sgr A* is slightly inverted with a
``submm-bump'' and a steep cut-off towards the IR
(e.g., Serabyn et al. 1997\nocite{SerabynCarlstromLay1997},
Falcke et al. 1998\nocite{FalckeGossMatsuo1998}), 
while the radio flux is variable on scales of weeks to months
(e.g., Wright \& Backer 1993, Falcke 1999b\nocite{WrightBacker1993,Falcke1999a}).  Sgr A* also shows an
unusually high ratio of circular to linear polarization
\cite{BowerFalckeBacker1999}.

The latest information comes from the detection of Sgr A* with the
X-ray satellite Chandra
\cite{BaganoffAngeliniBautz1999,BaganoffMaedaMorris2000}.  The first
epoch data show a point source at the location of Sgr A* with an X-ray
luminosity roughly two times below the earlier ROSAT limit
\cite{PredehlTruemper1994} and a photon index of $\sim$2.75$^{+1.25}_{-1.00}$.  This 
new measurement provides a crucial constraint for any model of
radiative emission from Sgr A*.

While all current models for Sgr A*'s radio emission consider
accretion onto the black hole as the driving force, they separate into
two generic classifications.  In models proposed by, e.g.,
Rees \cite*{Rees1982}, Melia \cite*{Melia1992a}, and
Narayan et al. \cite*{NarayanYiMahadevan1995}, the radio emission is produced by
processes in the accretion flow itself.  Alternatively,
Falcke et al. \cite*{FalckeMannheimBiermann1993}
propose that the radio emission stems from an outflow (see also
Reynolds \& McKee 1980\nocite{ReynoldsMcKee1980}) originating in
the accretion disk.

One currently popular accretion model is the Advection-Dominated
Accretion Flow (ADAF; Narayan et al. 1995\nocite{NarayanYiMahadevan1995}). While it may
explain the faintness of optical and ultra-violet (UV) emission, its
application to explain the compact radio emission from low-luminosity
AGN (LLAGN) is problematic.  For instance, in two well-known LLAGN
with radio nuclei similar to Sgr A*, M81
\cite{BietenholzBartelRupen2000} and NGC~4258
\cite{HerrnsteinGreenhillMoran1998}, intense observations with VLBI
revealed core-jet structures but found no evidence for radio emission
from an ADAF.  Furthermore, a systematic survey of LLAGN with the VLA
and the VLBA \cite{NagarFalckeWilson2000,FalckeNagarWilson2000} found
many compact radio nuclei, none of which show the highly inverted
spectrum expected in the ADAF model. On the other hand, the rather
flat radio spectra found are quite naturally explained within jet
models.  For Sgr A* itself, the standard ADAF model falls short of
explaining the cm-wave radio emission by more than an order of
magnitude and additional assumptions must be imposed in order to match
the spectrum
\cite{Mahadevan1998,OzelPsaltisNarayan2000}.

The difference in spectral index between a jet and an ADAF model stems
from the fact that in the latter the energy of the radiating particles
(i.e.~temperature) is a function of radius due to the dissipation of
energy in the viscous accretion flow, while an unperturbed supersonic
jet is to first order dissipationless and quasi-isothermal.  We mean
this in the sense that there is only limited cooling, due to adiabatic
losses which result from the longitudinal pressure gradient (see
Eq. \ref{jeteqn}).  Hence, the relative spectral indices reflect the
fundamentally different concepts underlying these models.

Given these issues, it would seem timely to revisit the jet model with
the additional constraints provided by the Chandra observations,
especially as Lo et al. \cite*{LoShenZhao1998} have claimed the first tentative
evidence for a jet structure in Sgr A* from 7 mm VLBI. Here, we
present a numerical calculation of the jet model using more realistic
electron distributions and the contribution of synchrotron
self-Compton (SSC) emission.  We note that the low X-ray fluxes
detected by Chandra were in fact expected in models invoking SSC
emission (e.g., Beckert \& Duschl 1997\nocite{BeckertDuschl1997}) and predicted for the jet
model \cite{Falcke1999b}.

\section{The Jet Model}
The basic model was already described in detail in
Falcke \cite*{Falcke1996a}. Symmetrically on either side of the accretion
flow a magnetized, relativistic proton and electron plasma (adiabatic
index $\Gamma=4/3$) is ejected from a nozzle where it becomes
supersonic, with an initial sound speed (in units of the speed of
light $c$) of $\beta_{\rm
s,0}=\sqrt{(\Gamma-1)/(\Gamma+1)}\sim0.4$. Each jet accelerates along
its axis through its pressure gradient and expands sideways with its
initial sound speed. The velocity field with bulk Lorenz factor
$\gamma_{\rm j}$ is then given by the Euler equation (see Eq.~1 in
Falcke 1996a\nocite{Falcke1996a})
\begin{equation}\label{euler3}\label{v} {\partial\gb\over\partial z}
\left({\left({\Gamma+\xi\over\Gamma-1}\right)(\gb)^2-\Gamma\over\gb}\right)={2\over z}
\end{equation}
with
$\xi=\left(\gb/(\Gamma(\Gamma-1)/(\Gamma+1))\right)^{1-\Gamma}$. As
before, the gravitational potential is ignored since its influence is
rather small in the supersonic regime considered here. The size of the
nozzle $z_0$, i.e.~the location of the sonic point, remains a free
parameter, since the exact launching mechanism for astrophysical jets
is unknown.

Given an initial magnetic field $B_0$, a relativistic electron total
number density $n_0$ with a characteristic electron energy
$\gamma_{\rm e,0}m_{\rm e}c^2$, the radius $r_0$ of the nozzle, and taking
only adiabatic cooling due to the longitudinal pressure gradient
(i.e.~$\propto {\cal M}^{\Gamma-1}$, where ${\cal M}$ is the Mach
number) and dilution by the lateral expansion into account, one can
determine the magnetic field $B(z_*)$, particle density $n(z_*)$,
electron Lorentz factor $\gamma_{\rm e,0}(z_*)$, and jet radius as a
function of the dimensionless distance from the nozzle
$z_*=(z-z_0)/r_0$:
\begin{eqnarray}\label{jeteqn}
&&{\cal M}(z_*)={\gb\over\gamma_{\rm s,0}\beta_{\rm s,0}},
 \; n(z_*)=n_0\cdot\left({r(z_*)/ r_0}\right)^{-2}{\cal M}^{-1}(z_*),\nonumber\\
&&r(z_*)=r_0+z_*/{\cal M}(z_*),\;\gamma_{\rm e}(z_*)=\gamma_{\rm e,0}\cdot{\cal M}^{-1/3}(z_*),\nonumber\\
&&B(z_*)=B_0\cdot\left({r(z_*)/r_0}\right)^{-1}{\cal M}^{-2/3}(z_*).
\end{eqnarray}
This fixes the basic parameters for synchrotron and SSC emission along
the entire jet.

By approximating the jets as a series of cylindrical sections, we can
calculate the total emission by integrating over the contributions
from each component.  For each segment the optical depth to
synchrotron absorption is $\tau_\nu=\frac{\pi}{2} \alpha_\nu
r(z_*)/D\sin{\theta_i}$, where $\theta_i$ is the angle between the jet
axis and the line of sight, $\alpha_\nu$ is the absorption
coefficient, $D=[\gamma(1-\beta\cos{\theta_i})]^{-1}$ is the Doppler
factor accounting for the angle aberration
(e.g., Lind \& Blandford 1985\nocite{LindBlandford1985}) due to the relativistic bulk
velocity, $\beta(z_*) c$, in the jet.  Using the transfer equation for
source-only emission, assumed constant within the segment,
$I_\nu(\tau_\nu)=(1-{\rm e}^{-\tau_\nu})S_\nu$ where
$S_\nu=j_\nu/\alpha$ is the source function.  Assuming isotropic
emission in the rest frame of the cylindrical shell, the net flux out
of the component is thus $F_\nu=4\pi I_\nu$.  The observed flux
density is then $F'_{\nu'}=D^2 2 r (D\sin{\theta_i}) \Delta z F_\nu/4
\pi d_{\rm gc}^2$, where the distance to the Galactic Center is assumed to
be $d_{\rm gc}=8.5$ kpc and the $2 r (D\sin{\theta_i})\Delta z$ factor is
the approximate projected surface area of the radiating cylinder.  In
the limits of $\tau\to0$ and $\tau\to\infty$ we recover the correct
optically thin and optically thick solutions.

To calculate the inverse-Compton up-scattered emission for the same
segment, we can ignore projection effects since the optical depth for
Compton scattering is small and use the self-absorbed synchrotron
emission in the component frame calculated above.  Then, using the
general expression for Compton scattering (e.g.,
Blumenthal \& Gould 1970\nocite{BlumenthalGould1970}) by a distribution of relativistic
electrons (including the Klein-Nishina limit), we find the SSC
emission in the frame of the component which then is transformed into
the observer's frame as before.

Relativistic beaming and possible -- but model-dependent and difficult
to quantify -- absorption in an accretion flow limit the visibility of
the second jet pointing away from Earth (see, e.g., the faint
counter-jet in the almost edge-on disk of NGC4258;
Herrnstein et al. 1998\nocite{HerrnsteinGreenhillMoran1998}). For simplicity the counter jet
is therefore ignored in our calculations. The most noticeable effect
of this jet for the range of parameters discussed here
($\theta_i\sim50^\circ$, $\gamma_{\rm j}\beta_{\rm j}=0.4-3$) could be a
$\le30\%$ increase in flux density at the highest frequencies from the
brighter nozzle component, quickly dropping to 10\% and less in the
mm-wave range. We point out that we also ignore here -- as with all
other current models for Sgr A* -- any general relativistic effects on
the spectrum and the emission region (see
Falcke et al. 2000a\nocite{FalckeMeliaAgol2000}). Thus, possible corrections due to the
counter-jet and the light propagation in the Kerr metric have to be
absorbed by the other parameters in the model. Given the overall
simplicity of the model we feel this is justified.

\section{Results}
\subsection{Radio Spectrum}

In theory, this model has six free parameters: the radius of the
nozzle, $r_0$, the location of the sonic point (corresponding to the
length of the nozzle),$z_0$, the magnetic field, $B_0$, the
equipartition factor, $k$ (which determines the electron density
$n_0$), the inclination angle, $\theta_i$, and the characteristic Lorentz
factor of the electron distribution, $\gamma_{\rm e,0}$.  In earlier
papers (e.g., Falcke et al. 1993\nocite{FalckeMannheimBiermann1993}) we used the jet-disk
coupling to determine some of these parameters, however, most
parameters are in fact well-constrained by data or obvious physical
arguments.

Firstly, the radius of the nozzle cannot be any smaller than the event
horizon of the black hole, while the above mentioned VLBI observations
provide an upper limit of $\la15$ Schwarzschild radii. The same is
true for the sonic point and nozzle height $z_0$ --- indeed the final
solution can be obtained with a symmetric nozzle,
i.e. $z_0\sim2r_0$. In addition, as shown in Falcke \cite*{Falcke1996b},
observed the high- and low-frequency turnover as well as the peak flux
of the submm-bump directly determine the magnetic field, electron
Lorentz factor, and density in the nozzle region subject only to
uncertainties in the measured spectrum.  Fitting the submm-bump with
the additional constraint of an equipartition solution, therefore
limits the number of entirely ``free'' parameters to one, i.e.~the
inclination angle $\theta_i$. Of course, the most likely inclination
angle for a randomly oriented source is around 57$^\circ$ and very
small or very large values for $\theta_i$ appear unlikely but are also
not completely excluded.

Furthermore, a steep cut-off or highly peaked electron distribution is
necessitated by the stringent IR upper limits.  Consequently, we see
no optically thin high-frequency emission as exists, for example, in
blazars, which would be due to a power-law tail of high-energy
electrons.

Here, we explore two possibilities for common astrophysical electron
distributions. First, we consider a narrow power-law with a standard
energy index $p=2$ and a sharp cut-off at roughly $5\gamma_{\rm e,0}$.
We also consider a relativistic thermal Maxwellian distribution with
$\gamma_{\rm e,0}\approx3.5\frac{kT}{m_{\rm e}c^2}$.  An alternative
energy distribution, i.e. one which is produced via hadronic
processes, will be explored in a later paper
\cite{MarkoffFalckeBiermann2000}.  However, the basic results remain
rather unchanged as long as the distribution is relatively narrow.

\begin{figure}
\centerline{\psfig{figure=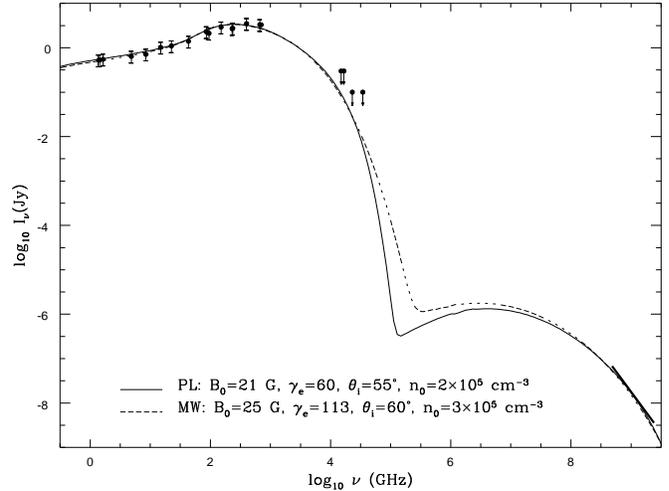,width=\figwidth,angle=-90}}
\caption[]{\label{sgrx-pl}Broad-band spectrum of Sgr A*. The dots are
the simultaneous spectrum measured by Falcke et al.~(1998) with
additional high-frequency data discussed by Serabyn et al.~(1997). We
have added 10\% errors for typical short-term variability.  This is
quite conservative---from a theorists point of view, as Sgr A* is
known to vary up to 30-100\% during flares
\cite{ZhaoBowerGoss2000,TsuboiMiyazakiTsutsumi1999}, and we have not
accounted for systematic errors in the mm-range.  In the hard X-rays
we show the possible detection of Sgr A* with Chandra in the range
2-10 keV which is relatively unaffected by absorption
\cite{BaganoffMaedaMorris2000}.  We show our model spectrum for a
power-law distribution of electrons (PL) and a relativistic Maxwellian
distribution (MW). The radius of the nozzle is $r_0\approx4R_{\rm s}$,
while its height is $z_0\approx2r_0$.}
\end{figure}

Fig. \ref{sgrx-pl} shows the best fits to the submm- and cm-wave
radio spectrum for the two electron distributions under the condition
of equipartition, with nozzle parameters given in the plots.  We show
$F_\nu$ rather than $\nu F_\nu$ because it is more conducive for
judging the quality of the spectral fit at cm-waves. The nozzle
component accounts for most of the submm-bump, as well as the main
Compton component, and the low-frequency spectrum stems from the
emission of the more distant parts along the jet. Within the model,
the slope of the cm-wave spectrum and the ratio between cm- and
submm-emission is mainly determined by the inclination angle. The
parameters we obtain for the jet and nozzle are close to those used by
Falcke \cite*{Falcke1996b},
Beckert \& Duschl \cite*{BeckertDuschl1997}, Falcke \& Biermann \cite*{FalckeBiermann1999}. 

\nocite{FalckeGossMatsuo1998}\nocite{SerabynCarlstromLay1997}

\begin{figure}
\centerline{\psfig{figure=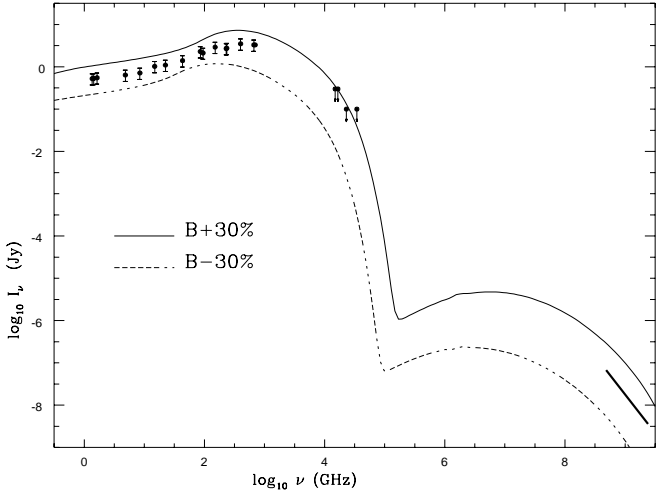,width=\figwidth,angle=-90}}
\caption[]{\label{bplot}Same as the power-law curve in Fig. \ref{sgrx-pl}, with
the magnetic field plotted for values 30\% higher (solid) and lower
(dashed) than what is used for the best fit.}
\end{figure}

\begin{figure}
\centerline{\psfig{figure=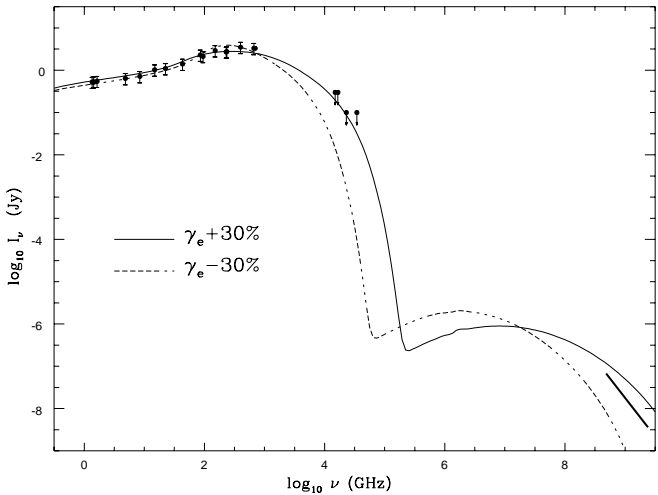,width=\figwidth,angle=-90}}
\caption[]{\label{gplot}Same as the power-law curve in Fig. \ref{sgrx-pl}, with
the characteristic electron Lorentz factor plotted for values 30\%
higher (solid) and lower (dashed) than what is used for the best fit.}
\end{figure}

\subsection{X-ray flux}
Once the radio spectrum due to synchrotron emission is fixed, so is
the Compton up-scattered spectrum. The peak frequency of the latter is
$\sim4\gamma_{\rm e,0}^2$ times the peak frequency of the synchrotron
emission being up-scattered. Hence, in order to produce soft X-ray
emission from 1 THz synchrotron emission one needs
$\gamma_{{\rm e},0}\sim100$.  It is interesting to note that the
characteristic $\gamma_{{\rm e},0}$ (either the cut-off, for the power-law,
or the peak, for the Maxwellian), discussed above in Sect. 3.1,
required to both fit the correct radio emission from the nozzle while
satisfying the IR upper limits falls exactly in this range.

To show how sensitive the model is to changes in parameters we display
the same version of the power-law model with the crucial fit
parameters $B_0$ and $\gamma_{\rm e,0}$ varied by 30\%
(Figs. \ref{bplot} \& \ref{gplot}). Changes in the magnetic field mainly reflect a change in
the total flux level while changes in the electron energy most
strongly affect the IR and X-ray emission. The Chandra detection
therefore limits the $\gamma_{\rm e,0}\la 130$, depending on the
electron distribution. Attempting to fit both the radio and X-rays
while remaining under the IR limits specifies a relatively tight range
in parameters.

However, it is not completely clear if the Chandra X-rays originate in
the jets or the accretion flow.  The lower flux compared to that
reported by ROSAT ($1-2\cdot10^{34}$ erg s$^{-1}$ within 0.8--2.5 keV,
see Predehl \& Zinnecker 1996\nocite{PredehlZinnecker1996}) suggests that SSC could be an
important component, however, thermal bremsstrahlung from the accretion
flow may still play an important role even though the spectrum appears
too steep at present.  The true test will be comparing the predictions
of the jet SSC component to future observations.  These predictions
are quite clear: we would expect significant, variable emission in the
EUV to X-ray band, with the variability correlated with the already
observed fluctuations in the submm-bump. The timescale for variations
in the radio are between two weeks and several months
(e.g., Falcke 1999b\nocite{Falcke1999a}). Typical strong but perhaps infrequent
flares in the mm-range seem to have timescales of about 20 days
\cite{ZhaoBowerGoss2000,TsuboiMiyazakiTsutsumi1999}. Similar
timescales are expected for the X-rays.

We also expect some curvature in the X-ray slope, rather than a pure
power-law.  Even if a mixture of jet SSC and emission from the hot
accretion flow account for the X-ray flux of Sgr A*, the nozzle should
still reveal itself through the correlated (with the radio) soft X-ray
emission.  Complete absence of X-ray variability would argue against
SSC emission giving a major contribution.

If we do not attempt to account for the X-rays via SSC, the model is
significantly less constrained, allowing for a larger range in the
X-ray output (as long as it is under the Chandra limits) and thus in
several of the parameters.  This could later be fitted to a value if
the assumedly (in this case) dominant disk and extended gas
contribution is specified.  However, if the Chandra flux proves to be
variable in correlation with the radio, then the SSC component is a
likely solution.

\subsection{VLBI Size and Extended Emission}
Possibly the most important constraints for any model are the VLBI
measurements of the size of Sgr A*. Since most models have a
stratified structure, the size of Sgr A* is expected to be a function
of frequency.  In our model, for a given observing
frequency, the emission is highly concentrated to one
spatial scale and thus predicts very little extended emission beyond the
core (which is basically the $\tau=1$ surface of the radio jet).  The
emission from the jets at a particular frequency is self-absorbed at
small distances from the origin and cut off at large distances where
the decreased magnetic field shifts the synchrotron cut-off frequency
below the observing frequency.  Thus, extended emission from the most
visible jet is highly suppressed and the size of the detectable core
will be a power-law $z\propto\nu^{-m}$ with $m\sim0.9-1$. This also
implies a shift of the location of the core with frequency. 

\begin{figure}
\centerline{\psfig{figure=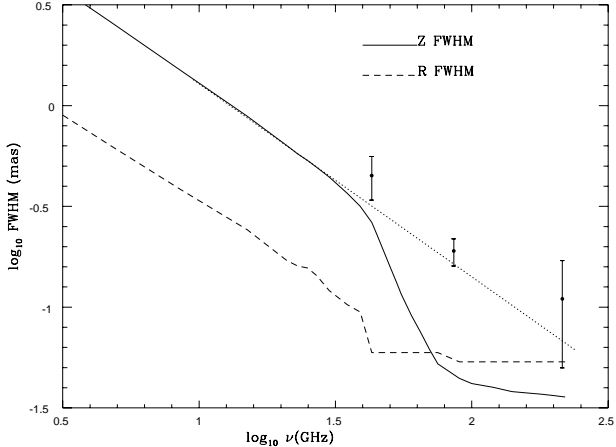,width=\figwidth,angle=-90}}
\caption[]{\label{sgrx-size}Projected FWHM of the major and minor axis
of the dominant jet as a function of frequency. The filled dots mark
the FWHM as measured by Lo et al.~(1998; 43 GHz) and Krichbaum et
al.~(1998; 86 \& 215 GHz). At frequencies above 30 GHz one obtains a
two component structure with an increasingly stronger core (nozzle,
solid dashed line) and a fainter jet component (dotted line).}
\end{figure}

Fig. \ref{sgrx-size} compares the projected full width at half
maximum (FWHM) of major and minor axis of the emission predicted by
the jet model (for the jet pointing towards us) together with the
upper limits imposed by high-frequency VLBI observations. Given the
calibration difficulties (see discussion in
Bower et al. 1999b\nocite{BowerFalckeBacker1999b}), the values plotted can also be
considered as upper limits on the source size in {\it one} direction
(measurements at 3 and 1.4 mm wavelengths cover only a very restricted
range in the (u,v)-space -- see Doeleman et al. 1999\nocite{DoelemanRogersBacker1999} for a
nice discussion of this problem).

Throughout the cm-wave range the predicted emission basically
resembles one elliptical component decreasing in size with wavelength
and only at mm-waves (i.e.~above 30 GHz) does the more compact
core-component, the nozzle, appear together with the more extended
jet-emission. Comparing data and model in Fig.\ref{sgrx-size} shows
the extended jet emission is within the constraints imposed by
VLBI---component ejections or a faint counter-jet could always make the
jet somewhat larger. The exact size is also a function of the current
flux density of Sgr A* and hence should vary on the same timescales
found in the spectrum (see Falcke \& Biermann 1999\nocite{FalckeBiermann1999} for an analytic
estimate of the functional dependence).

\nocite{LoShenZhao1998,KrichbaumGrahamWitzel1998}

\section{Discussion and Summary}

We have calculated the emission expected from a jet model in the
context of the Galactic Center. The spectrum of Sgr A*, including the
new X-ray observations from Chandra, can be modeled entirely by
emission from this jet. We can also show that the radio emission
satisfies all constraints imposed by VLBI observations. This indicates
that the basic jet model introduced by
Falcke et al. \cite*{FalckeMannheimBiermann1993} can provide a detailed explanation
of the Sgr A* radio and X-ray spectrum.

An important feature of the model is the highly peaked electron
distribution (see also Beckert \& Duschl 1997\nocite{BeckertDuschl1997}) which has a number
of interesting implications:
\begin{itemize}
\item[a)] In typical AGN core-jet sources the extended jet structure is due to
an optically thin power-law. Here this extended emission is greatly
suppressed due to the steep cut-off in the electron spectrum, required
by the IR limits. This naturally can explain the compact jet structure
as seen by Lo et al. \cite*{LoShenZhao1998}. 
\item[b)] The X-ray emission produced via up-scattering of the
synchrotron photons (SSC) can have a rather steep spectrum which is
not a perfect power-law. This can well explain the very soft emission
found for Sgr A* and perhaps also for M31*
\cite{GarciaMurrayPrimini2000} that otherwise seems to be too steep
for thermal emission from accretion flows.
\item[c)] The nature of this distribution with $\gamma_{\rm e}\la100$
requires a closer look. The characteristic energy derived here for the
electrons (or positrons) could be indicative of hadronic processes as
discussed in Falcke \cite*{Falcke1996b}, Mahadevan \cite*{Mahadevan1998}, and
Markoff et al. \cite*{MarkoffFalckeBiermann2000}.
\item[d)] Finally, in light of recent puzzling polarization
measurements
\cite{BowerBackerZhao1999,BowerFalckeBacker1999,BowerWrightBacker1999}, 
one needs to reconsider the polarization properties of synchrotron
sources with such unusual electron distributions.
\end{itemize}

Our primary assumption is the existence of a nozzle close to the
central black hole, which collimates a relativistic plasma having
equipartition between the magnetic field and particles.  The
requirement that this nozzle is responsible for the dominant
submm-bump in the radio, and possibly the SSC X-ray emission, allows
us to fix most free parameters.  Once the parameters for the nozzle
are fixed by fitting the submm-bump, the evolution of magnetic field
and density along the jet does not require additional parameters.  The
spectra we obtain are therefore generic for collimated outflows from
any accretion flow---whether a magneto-hydrodynamical jet from a
standard accretion disk or an outflow from an ADAF---provided the
accretion flow can produce the required magnetic field, electron
temperature, and density near its inner edge. For an ADAF or
Bondi-Hoyle type accretion, the inclusion of jets near the black hole
could thus enhance those models in accounting for the cm-wave radio
emission. The energy requirements to produce such jets (see
Falcke \& Biermann 1999\nocite{FalckeBiermann1999}) are rather small compared to the power
available through accretion of nearby winds
\cite{CokerMelia1997}. Therefore, it is possible that most of the
visible emission in Sgr A* could be produced by an AGN-like jet and
not an accretion flow. 

Because the model is so generic, it is possible that this model finds
an application also in other low-luminosity AGN or even X-ray
binaries. Here it needs to be checked whether in some cases similarly
peaked electron distributions are present and whether some soft X-ray
emission could be related to SSC emission from the jet.

\begin{acknowledgements}
We thank Mark Morris and Frederick K. Baganoff for providing us with
information about the Chandra observations of Sgr A* before
publication.
\end{acknowledgements}

\bibliography{aamnem99,ref0}
\bibliographystyle{aabib99}

%\newpage % if processed with CM fonts, balances the references columns
\clearpage

\end{document}